\input harvmac
\overfullrule=0pt

\Title{\vbox{\baselineskip12pt
\hbox{hep-th/0205192}
\hbox{OHSTPY-HEP-T-02-016} }}
{\vbox{\centerline{A proposal to resolve the black hole
  }{\centerline{ information paradox${}^*$}}}}

\smallskip
\smallskip
\centerline{\bf Samir D. Mathur}
\smallskip
\bigskip

\centerline{\it Department of Physics}
\centerline{\it The Ohio State University}
\centerline{\it Columbus, OH 43210, USA}
\medskip
\centerline{mathur@mps.ohio-state.edu}
\bigskip

\medskip

\noindent

The entropy and information puzzles arising from  black holes
cannot be resolved if quantum gravity effects remain confined to
a microscopic scale. We use  concrete computations in
nonperturbative string theory to argue for three kinds of  nonlocal
effects that operate over macroscopic distances. These effects arise 
when we make a  bound state of a
large number of branes, and occur at the
correct scale to resolve the paradoxes associated with black holes.

\vskip 2.0 true in

*This essay received an ``honorable mention'' in the
	Annual Essay Competition of the Gravity Research
	Foundation for the year 2002.

\Date{}

Black
holes have a `Bekenstein entropy' $S_{Bek}=A/4G$. Are there 
$e^{S_{Bek}}$ microstates
of the hole, and if so,  then where do we  see the
differences between them? The Hawking radiation from the hole
appears to be reliably given by a semiclassical calculation, but
the radiation thus computed carries little information about the
infalling matter.  Is    unitarity  violated by black holes?

  The essential
strength of these puzzles lies in the fact that the more massive
we make the hole, the smoother its geometry becomes at the
horizon. Incoming matter disappears
into the central singularity while Hawking radiation  arises from 
vacuum fluctuations
near the smooth horizon. But the horizon is a large distance --
$R_s$, the Schwarzschild radius -- from the singularity, so the
radiation is unable to encode the information of the infalling
matter.

While everyone agrees that nonlocal physics can arise at a
microscopic length scale like planck length or string length, what
we need to effect the information transfer is nonlocal effects
across spatial lengths of order $R_s$, which is a length that {\it
increases} with the mass of the hole. Thus in our theory of
gravity if we make a bound state of a large number of quanta, then
nonlocal effects should operate not at a fixed microscopic length
but rather at a length that increases with the number of quanta in
the bound state.

In this article we argue that we must modify our intuition about 
quantum gravity to include three phenomena where
quantum gravity effects indeed reach out to macroscopic length scales:

  (a) \quad Bound
states of branes have a {\it nonzero size}; it is important that
  this size {\it grows} with the number of branes.

(b) \quad When we make a bound state from
  a large number $N$ of branes  then
the excitations of this bound state are given in terms of {\it
fractionated} branes which have a tension $\sim 1/N$; thus the
virtual fluctuations of these fractionated branes extend to
macroscopic distances when $N$ is macroscopic.

(c)  \quad The concept of
{\it limited stretchability}, which says that if
spatial hypersurfaces `stretch too much' during evolution then we encounter a
breakdown of the semiclassical approximation even though there are
no large curvatures anywhere -- rather,  the degrees of freedom on
the hypersurface have become too `dilute'.

We show that these
principles are supported by concrete calculations in string
theory, and that they fit together to provide a resolution of the
paradoxes arising from black holes.\foot{For some other attempts at 
resolving the information puzzle (both related and unrelated)
see \ref\all{E.~Verlinde and H.~Verlinde,
Nucl.\ Phys.\ B {\bf 406}, 43 (1993)
[arXiv:hep-th/9302022]; Y.~Kiem, H.~Verlinde and E.~Verlinde,
Phys.\ Rev.\ D {\bf 52}, 7053 (1995)
[arXiv:hep-th/9502074]; L.~Susskind,
Phys.\ Rev.\ D {\bf 49}, 6606 (1994)
[arXiv:hep-th/9308139]; L.~Susskind and P.~Griffin,
arXiv:hep-ph/9410306; D.~A.~Lowe and L.~Thorlacius,
Phys.\ Rev.\ D {\bf 60}, 104012 (1999)
[arXiv:hep-th/9903237]; H.~S.~Nastase,
quantum gravity unified with other interactions,''
arXiv:hep-th/9601042.  Several additional references can be found in 
S.~B.~Giddings,
arXiv:hep-th/9508151.}.}

We will extract evidence for these  principles from computations
in IIB string theory, which lives in $9+1$ spacetime dimensions.
Five space directions are compactified; we write these as
$T^4\times S^1$. One system of interest is the D1-D5 system -- we
wrap $n_5$ D5 branes on $T^4\times S^1$, and $n_1$ D1 branes on
$S^1$. The bound state of these branes has a highly degenerate
ground state, with entropy $S_{micro}=2\sqrt{2}\sqrt{n_1n_5}$. The 
volume of $T^4$ is $(2\pi)^4
V$, the radius of  $S^1$ is $R$, the string coupling is $g$,
and the string tension is $\alpha'$.  A set of dualities maps this
`2-charge system' to another 2-charge system -- the FP bound
state, where we have $n_5$ fundamental strings wrapped on
$S^1$, carrying $n_1$ units of momentum along  $S^1$.

Let us first look at the size of bound states.
Consider a fundamental string (F) carrying momentum (P). The string 
possesses no longitudinal
vibration modes, so  to carry the momentum the string must bend away 
from its central axis. The FP state thus
acquires a certain transverse size;  the more the momentum, the 
larger this size.

Had the bound state been pointlike, the metric would have ended in a 
point singularity at $r=0$. But the finite size of the bound
state modifies this naive metric inside a region $r<r_0$. A recent 
analysis of such metrics yielded an interesting set of results.
For the same total energy the string can carry different vibration 
profiles, and these give metrics that differ for $r<r_0$.
\ref\lmone{
O.~Lunin and S.~D.~Mathur,
Nucl.\ Phys.\ B {\bf 610}, 49 (2001)
[arXiv:hep-th/0105136].}\ref\lmtwo{
O.~Lunin and S.~D.~Mathur,
Nucl.\ Phys.\ B {\bf 623}, 342 (2002)
[arXiv:hep-th/0109154].}.   Let us coarse-grain over these 
`microstates' by truncating  the geometry at the surface
$r=r_0$.  From the area $A$ of this `horizon' we
find \ref\lmthree{
O.~Lunin and S.~D.~Mathur,
arXiv:hep-th/0202072.}
\eqn\one{S_{Bek}\equiv A/4G\sim \sqrt{n_1n_5}\sim S_{micro}~!}
Further, a
quantum that falls past this `horizon' gets `trapped' in the 
complicated geometry at $r<r_0$, for times
  $\sim 1/\hbar$.  So from a classical perspective, this surface 
indeed behaves like a horizon.

Thus for the  extremal 2-charge system we resolve the `entropy 
puzzle': we directly relate the `area entropy' to a coarse graining
over microstates (`hair'). Tracing back, the critical fact was that 
the bound states had  a
transverse size which grew with the charges at just the correct rate 
to always reach a  `horizon' which satisfies
\one .

Can such a picture can be extended to the
3-charge system, which can be constructed by adding $n_p$ units of
momentum along $S^1$ to the D1-D5 bound state? The 3-charge system
has a classical size horizon radius, so  we  need to understand how
the microstate develops an effective size that is classical. The
key notion (for both two and three charge states) is  {\it 
fractionation}.  Consider a string
wrapped on a circle of radius $R$.  The minimum
excitation energy (for no net momentum along the string) is  ${1\over 
R}+{1\over R}={2\over R}$. But now let the
string wind $N$ times around the circle before closing on itself.
Now the vibration modes  can have a wavelength $2\pi
R N$, and the energy gap drops to ${2\over NR}$ \ref\dasmathur{
S.~R.~Das and S.~D.~Mathur,
Phys.\ Lett.\ B {\bf 375}, 103 (1996)
[arXiv:hep-th/9601152].}.

If we have a bound state of D1 and D5 branes then for
  vibrations along $S^1$  the energy spacing drops to
${2\over n_1n_5 R}$ \ref\maldasuss{
J.~M.~Maldacena and L.~Susskind,
Nucl.\ Phys.\ B {\bf 475}, 679 (1996)
[arXiv:hep-th/9604042].}.  Clearly, the larger we  make the hole, the 
lighter will be its excitations. In
\ref\mathurone{ S.~D.~Mathur,
Nucl.\ Phys.\ B {\bf 529}, 295 (1998)
[arXiv:hep-th/9706151].} the
effective size of the D1-D5-momentum bound state resulting from 
`fractionated' degrees of freedom was estimated at
weak coupling, and was found to be
$$r_0\sim  [{(n_1n_5n_5)^{1/2}g^2\alpha'^4\over VR}]^{1/3}$$
But this is the same algebraic function of these six parameters
that gives (upto a factor of order unity) the Schwarzschild radius
of the 3-charge hole! These results suggest that
fractionation gives  bound states of branes a size that always equals 
the horizon radius,  We picture a `fuzz' of virtual
brane-antibrane pairs extending out to the horizon. A massive 
infalling particle falls
through this fuzz to the center, but over a longer time scale
(order $1/\hbar$) these virtual fluctuations can transport the
information in the particle to horizon  and into the
outgoing Hawking radiation.

Let us now address the information paradox. Broadly speaking, we
can consider two kinds of foliations in analyzing the radiation.
In the first kind, slices are $t={\rm constant}$ for $r>>2m$, but 
inside the horizon they extend into the singularity.
Can the singularity at $r=0$ modify radiation at $r=2M$? Yes, since 
fractionation generates effects
that extend to the horizon!

But an apparently  stronger argument for information loss can be
made by using a foliation where the slices never approach the
singularity, and there are no large curvatures anywhere.
It was shown in \ref\mathures{
S.~D.~Mathur,
Int.\ J.\ Mod.\ Phys.\ A {\bf 15}, 4877 (2000)
[arXiv:gr-qc/0007011].} that in such foliations the spacelike slices
{\it stretch} in the course of evolution by a large factor $\sim 
1/\hbar$. It was suggested that
spacelike hypersurfaces should be regarded as `rubber sheets' that 
have not only a shape but a  density of degrees of freedom.
Stretching dilutes these degrees, and if more matter data is placed 
on the slice than the number of available `bits'
then nonlocal effects occur.
Quantitatively, if we assume that a volume $V$ in flat space can 
stretch to a `throat' with maximum depth
$\sim V/G$ then the information paradox is resolved -- every 
foliation has slices that  have  a singularity or `stretch
too much', and the semiclassical derivation of Hawking radiation is 
invalidated in each case.

Remarkably, a similar picture emerged in the exact analysis of throat 
geometries of the D1-D5 system, which was mapped by
AdS/CFT duality to a 1+1 sigma model \lmtwo. The throat had different 
depths for different microstates. The dual sigma
model described the geometry in terms of a collection of `component 
strings'. Longer throats (`stretched geometries') were
described in the CFT by fewer component strings. When enough quanta 
were placed in the throat so that  each component string
in the dual description was excited (all `bits' were used up),  the 
physics underwent a qualitative change. Lastly, the maximum
depth of the throat was exactly
$V/2G$ !

To summarize, fractionation generates very light degrees of freedom; 
the virtual fluctuations of these degrees can transport the
information in an infalling particle all the way to the horizon (over 
times of order $1/\hbar$). If we consider a foliation
by `regular slices' that avoid the singular source of these virtual 
quanta then the slices  `stretch' by a factor  $O(1/\hbar)$.
AdS/CFT duality supports the idea that degrees of freedom along the 
slice are limited and get `overdiluted' under such
extreme stretching, so  that semiclassical behavior breaks down. 
Physics in the absence of black holes, studied on `regular'
foliations, remains unaffected by all these phenomena, and  we 
resolve all puzzles associated with black holes.

\vfill
\eject
\listrefs

\end